\documentclass[aps,prb,reprint,floatfix,superscriptaddress]{revtex4-2}

\usepackage[english]{babel} 
\usepackage{amssymb,amsmath}
\usepackage{graphicx}
\usepackage{float}
\usepackage{url}
\usepackage{multirow}
\usepackage{chemformula}
\usepackage{tabularx}

\newlength{\smallpic}
\begin{document}

\title{Selenium and the role of defects for photovoltaic applications}

\author{Hadeel Moustafa} 
\affiliation{CAMD, Computational Atomic-Scale Materials Design, Department of Physics, Technical University of Denmark, 2800 Kgs. Lyngby Denmark}
\author{Jiban Kangsabanik}
\affiliation{CAMD, Computational Atomic-Scale Materials Design, Department of Physics, Technical University of Denmark, 2800 Kgs. Lyngby Denmark}
\author{Fabian Bertoldo} 
\affiliation{CAMD, Computational Atomic-Scale Materials Design, Department of Physics, Technical University of Denmark, 2800 Kgs. Lyngby Denmark}
\author{Simone Manti}
\affiliation{INFN, Laboratori Nazionali di Frascati, Via E. Fermi 54, I-00044 Roma, Italy}
\author{Kristian S. Thygesen} 
\affiliation{CAMD, Computational Atomic-Scale Materials Design, Department of Physics, Technical University of Denmark, 2800 Kgs. Lyngby Denmark}
\author{Karsten W. Jacobsen} 
\affiliation{CAMD, Computational Atomic-Scale Materials Design, Department of Physics, Technical University of Denmark, 2800 Kgs. Lyngby Denmark}
\author{Thomas Olsen} 
\email{tolsen@fysik.dtu.dk}
\affiliation{CAMD, Computational Atomic-Scale Materials Design, Department of Physics, Technical University of Denmark, 2800 Kgs. Lyngby Denmark}

\begin{abstract}
We present first principles calculations of the electronic properties of trigonal selenium with emphasis on photovoltaic applications. The band gap and optical absorption spectrum of pristine selenium is calculated from many-body perturbation theory yielding excellent agreement with experiments. We then investigate the role of intrinsic as well as extrinsic defects and estimate the equilibrium concentrations resulting from realistic synthesis conditions. The intrinsic defects are dominated by vacancies 
and we show that these do not result in significant non-radiative recombination. The charge balance remains dominated by vacancies when extrinsic defects are included, but these may give rise to sizable non-radiative recombination rates, which could severely limit the performance of selenium based solar cells. Our results thus imply that the pollution by external elements is a decisive factor for the photovoltaic efficiency, which will be of crucial importance when considering synthesis conditions for any type of device engineering.
\end{abstract}

\maketitle

\section{Introduction} 
In 1873, selenium was discovered to exhibit photoconductivity \cite{smith1873action}, which led to the creation of the first PV solar cells with an efficiency of less than 1\% \cite{ferry2020challenges} and by 1985 the record conversion efficiency of selenium had increased to 5\% \cite{nakada1985polycrystalline}. However, having a band gap of 1.85 eV \cite{madelung2004semiconductors, todorov2017ultrathin, john2017electronic}, selenium is far from being an optimal material for direct conversion of solar energy \cite{bishop2017record}, and silicon has by now become the dominant semiconductor material in the solar cell industry. Nevertheless, selenium has recently regained attention for usage in tandem solar cells where the efficiency may exceed the Shockley-Queisser limit for single juction solar cells \cite{shockley1961detailed}. Such devices require a small gap ($\sim1.0$ eV) material and a large gap ($\sim1.8-1.9$ eV) material. Since silicon already dominates the solar industry and has a 1.12 eV band gap, it is the best contender for the small gap material \cite{polman2016photovoltaic}, but the quest to find an optimal large gap material is still open. In addition to having a nearly optimal band gap for tandem cells \cite{madelung2004semiconductors, todorov2017ultrathin, john2017electronic}, selenium offers a  number of advantages in this regard. It is easily integrable with silicon devices, has a low production temperature and it maintains stability under humidity and in the presence of oxygen. 

There has recently been taken a number of important steps towards integrating selenium in photovoltaic devices. In Ref. \cite{todorov2017ultrathin} a record efficiency of 6.5\% was obtained by completely redesigning the single junction selenium device architecture. Subsequently, it has been shown that bifacial (required for tandem cells) selenium single-junction devices could be constructed with a power conversion efficiency of 5.2\% \cite{youngman2021semitransparent}. There is, however, a number of issues that need to be solved before selenium can become a successful partner in tandem solar cells. Most importantly, the efficiency is a factor of four lower than the theoretical limit for a 1.85 eV band gap material, implying significant losses. The origin of reduced efficiency was scrutinized in Ref. \cite{Nielsen2022} where it was concluded that non-radiative recombination is likely to play a crucial role and the optimization of the bulk photoconductive properties thus seems much more pertinent than optimizing device architectures.

In the present work we have applied first principles calculations to quantify the rate of non-radiative recombination from intrinsic defects. We find that the intrinsic defects are completely dominated by vacancies and while these govern the charge balance and carrier densities they do not give rise to non-radiative decay. This implies that the loss may originate from recombination at extrinsic defects and spurs hope that photoconversion efficiencies may be significantly increased by employing selenium growth conditions where the detrimental elements are absent. 

The paper is organized as follows. In Sec \ref{sec:computational} we provide the computational details applied in the present work and in Sec. \ref{sec:results} we present our results. We start by a thorough investigation of the electronic properties of pristine selenium and then characterize the properties of defects with respect to formation energies and induced carrier concentrations, which allow us to calculate the efficiency of energy conversion. In Sec \ref{sec:conclusion} we provide a conclusion and outlook.
\begin{figure}
\centering
\includegraphics[width=0.95\linewidth]{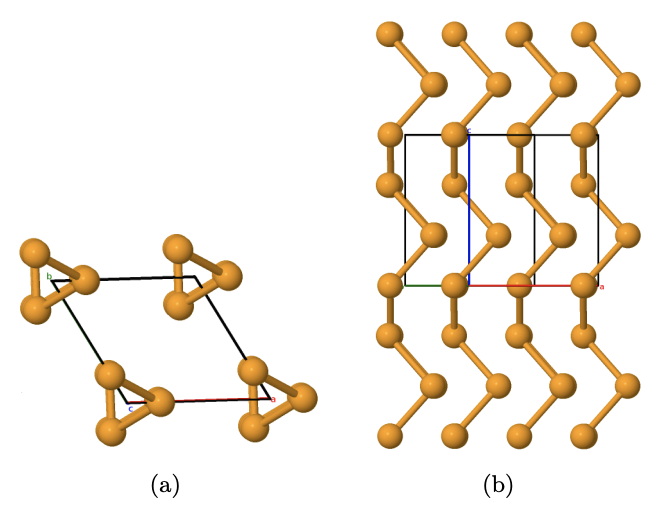}
\caption{Crystal structure of trigonal selenium viewed along (left) and transverse (right) to the selenium chains.}
\label{fig:selenium}
\end{figure}

\section{Computational details}\label{sec:computational}
All calculations were carried out with the electronic structure package GPAW \cite{enkovaara2010electronic}, which employs the projector augmented wave formalism \cite{blochl, kresse1999ultrasoft} and a plane wave basis set. In addition, we used the atomic Simulation Environment (ASE) \cite{Larsen2017} and the Atomic Simulation Recipes (ASR) \cite{Gjerding.2021}, which is a set of python modules that facilitate the workflows for defect calculations. Most of our calculations are based on density functional theory (DFT) and for these we applied a plane wave cutoff of 800 eV and a Fermi-Dirac smearing of 0.05 eV. The experimental lattice parameters were used for all calculations and the Kohn-Sham band gaps of the pristine system was obtained using a $60\times60\times60$ $\Gamma$-centered $k$-point grid using the PBE \cite{Perdew:1996ug}, SCAN \cite{sun2015strongly} and mBEEF \cite{wellendorff2014mbeef} functionals. Spin-orbit coupling was included non-selfconsistently \cite{Olsen2016a} in the band gap and band structure calculations. The inverse of the effective mass tensors for holes and electrons were computed by fitting second order polynomials to the valence band maximum and conduction band minimum.

In order to obtain a more accurate estimate for the band gap we also applied non-selfconsistent HSE06 \cite{heyd2003hybrid} and G$_0$W$_0$ \cite{huser2013quasiparticle} calculations using $k$-point grids of 13$\times$13$\times$11 and 12$\times$12$\times$10 respectively. For the G$_0$W$_0$ calculations we used full frequency integration, a plane wave cutoff of 200 eV and 568 bands for the screened interaction. The difference to the PBE Kohn-Sham eigenvalues were then interpolated on the $60\times60\times60$ grid and used to estimate the HSE06 and G$_0$W$_0$ band gaps. With both methods we find that the resulting band structures are rather accurately represented by applying a rigid shift between the conduction bands and valence bands.

The optical absorption spectrum was computed by solving the Bethe-Salpeter equation (BSE) as well as from the random phase approximation (RPA) \cite{Onida2002}. In both approaches, the PBE eigenstates and eigenvalues were used and the conduction bands were rigidly shifted to match the calculated G$_0$W$_0$ gap. The resulting spectrum was then rescaled to conserve the optical sum rule. The BSE calculations were carried out in the Tamm-Dancoff approximation with six valence bands, six conduction bands, and a $18\times18\times14$ $k$-point grid. We included 54 bands in the calculation of the screened interactions and a plane wave cutoff of 50 eV.

We have calculated formation energies of Se vacancies (V\textsubscript{Se}), interstitial defects (i\textsubscript{Se}) and the substitutional defects C, Si, Ge, Sn, N, P, As, Sb, O, S, Te, F, Cl, Br and I. The defect calculations were carried out with 3x3x3 (81 atoms) super cells, each having one defect and the structures were relaxed (with fixed lattice constants) until all forces on atoms were below 10 meV/{\AA}. We used a $k$-point grid of 4$\times$4$\times$4 and for each type of defect, we considered charge states of $q$=1, 0, -1, and -2 in units of $|e|$. The HSE06 functional yields the best agreement with the experimental band gap, but in order to perform a large number of calculations with different extrinsic defects in several charge states we have chosen to apply the mBEEF functional for the defect calculations in this work. A comparison between formation energies and density of states obtained with PBE, SCAN and mBEEF is presented in the supplemental material (SM) \cite{SM}  (see also Refs \cite{PhysRevB.104.064105,passler1976relationships,tiedje1984limiting} therein) where we also show the formation energies and carrier recombination rates obtained with HSE06 for the selenium vacancy. Finally, due to the long range nature of Coulomb interactions the charged defect calculations need to be corrected for interactions between periodic images, and we have applied the correction scheme of Ref. \cite{freysoldt2011electrostatic}, where the convergence with respect to super cell size is accelerated by correcting with a model charge distribution. The correction depends on the dielectric constant, which was calculated in the RPA. The formation energy of a defect $\mathrm{X}^q$ is then written as
\begin{equation}\label{eq:eform}
E^{f}\left[\mathrm{X}^q\right] = E_{\mathrm{tot}}\left[\mathrm{X}^q\right] - E_{\mathrm{tot}}^\mathrm{bulk} - \sum_i n_{i}\mu_i + qE_\mathrm{F} + \Delta_\mathrm{corr},
\end{equation}
where $E_{\mathrm{tot}}\left[\mathrm{X}^q\right]$ denotes the total energy of the relaxed structure with the defect $\mathrm{X}^q$ and $E_{\mathrm{tot}}^\mathrm{bulk}$ is the total energy of pristine selenium. The reference chemical potential of the atom species $i$ is denoted by $\mu_i$, and $n_i$ is the number of such atoms added ($n_i > 0$) or removed ($n_i <0$) to form the defect. $E_\mathrm{F}$ is the chemical potential of the electrons which we reference to the top of the valence band in pristine selenium.  $\Delta_\mathrm{corr}$ is the correction for the spurious interactions between localized charge states in periodic repetitions of the super cell as well as the interaction with a compensating homogeneous background charge. The charge transition level between $q$ and $q\pm1$ is determined as the value of $E_\mathrm{F}$ where $E^f\left[\mathrm{X}^q\right]=E^f\left[\mathrm{X}^{q\pm1}\right]$.
\begin{figure*}
    \centering
   \includegraphics[width=\textwidth]{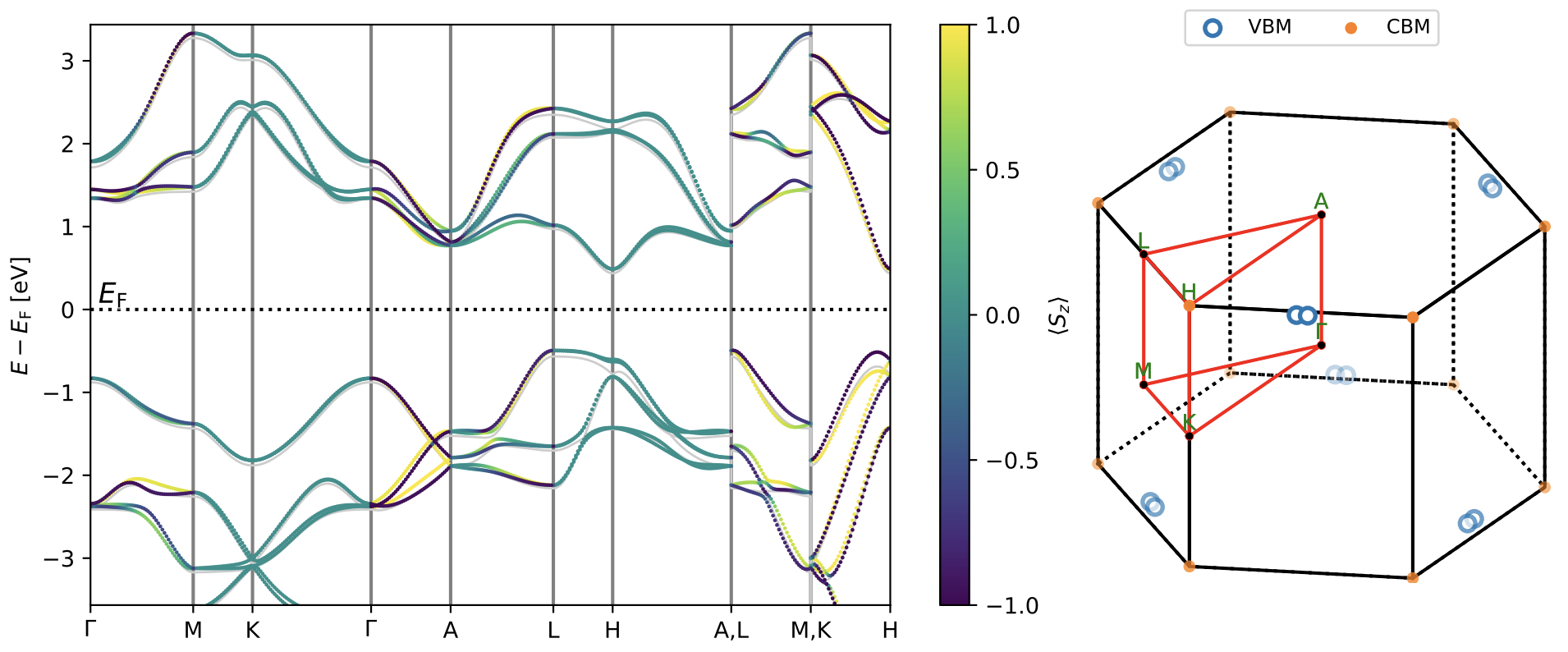}
    \caption{Left: PBE band structure of Se with (colored) and without (grey) spin-orbit coupling. The colors show the expectation values of $S_z$ for  the Kohn-Sham orbitals. Right: Brillouin zone of trigonal Se showing the path applied in the band structure as well as the positions of the conduction band minima (CBM) and valence band maxima (VBM).}
    \label{fig:BS}
\end{figure*}

\section{Results}\label{sec:results}
\subsection{Pristine selenium}
The trigonal phase of bulk selenium exhibits chiral chains of Se atoms that are bound by dispersive interactions and arranged in a hexagonal lattice. The chains are characterized by a three-fold screw axis and the primitive unit cell thus contains three Se atoms as shown in Fig. \ref{fig:selenium}. In addition, the structure has a two-fold axis of rotation orthogonal to the chains and the space group is thus P$3_{1}12$ (number 152).  There are several reports in the literature of additional selenium phases ($\alpha$-selenium \cite{cherin1972refinement}, $\beta$-selenium  \cite{marsh1953crystal}, monoclinic $\gamma$-selenium \cite{foss1980crystal} and other phases created by applying high pressure \cite{keller1977effect}), but the trigonal form is the only phase that is thermodynamically stable at room temperature. The lattice parameters of the hexagonal unit cell are $a=4.366$ {\AA} and $c=4.954$ {\AA} \cite{cherin1967crystal}. Due to the lack of inversion symmetry, trigonal selenium exhibits Weyl nodes, Rashba-type band splittings with non-trivial spin textures and has been shown to transit to a Weyl semimetal under strain \cite{PhysRevLett.114.206401}. The optical band gap is 1.95 eV \cite{Tutihasi1967, Stuke1970} and photoluminescence measurements have shown that there is an indirect gap situated at 1.80-1.85 eV \cite{ZETSCHE19691425, PhysRevLett.42.264}.

In Fig. \ref{fig:BS} we show the PBE band structure of pristine selenium with and without spin-orbit coupling. Due to the lack of inversion symmetry the spin degeneracy is lifted by spin-orbit coupling everywhere in the Brillouin zone except for the high symmetry points and certain high symmetry lines. The spin splitting is largest in the vicinity of the H point where the highest valence band is split by 0.23 eV. The $\Gamma$-A and K-H lines are invariant under the three-fold screw rotation (along the $z$-direction) and the spin projection along these lines are thus pinned along $z$. In addition, the $\Gamma$-K, M-K, A-H and H-L lines are invariant under various two-fold rotations (rotation axes perpendicular to $z$) and thus have vanishing spin projections along $z$. The spin structure and rashba-type splittings of trigonal selenium was analyzed in detail in Ref. \cite{PhysRevLett.114.206401} and will not be discussed in more detail here.

The conduction band minimum (CBM) is located at H and exhibits a small Rashba-type splitting when spin-orbit coupling is included. The valence band maximum (VBM) is located exactly at L when spin-orbit coupling is neglected, but shifts away from the high symmetry lines when spin-orbit coupling is included. This was also observed by Hirayama et al. \cite{hirayama2015weyl}. Due to the combination of two-fold rotational symmetry and time reversal symmetry, the VBMs constitute pairs that are symmetric under reflection in the $\Gamma$AM plane as shown in Fig. \ref{fig:BS}b and there are six such pairs due to a combination of the two-fold and three-fold axes. In addition to the global maximum in the vicinity of L there is a local maximum on the HK line, which is merely 25 meV lower than the VBM (0.2 eV lower without spin-orbit coupling). The indirect gap is 0.98 eV whereas the direct gap is 1.06 eV and is located at the H point. 
\begin{table}[b]
    \centering
    \begin{tabular}{ |l|r|r|r|} 
\hline
 & {\bf Band gap} & {\bf Direct gap} \\
\hline
PBE & 0.99 & 1.09 \\
\hline
HSE06@PBE & 1.72 & 1.82 \\
\hline
G$_0$W$_0$@PBE & 1.78 & 1.94 \\
\hline
SCAN &  1.25&1.37\\
\hline
mBEEF &1.43&1.55 \\
\hline
\end{tabular}
    \caption{The direct and indirect band gaps obtained from different functionals and $G_0W_0$. All gaps include spin-orbit coupling.}\label{tab:band_gaps}
\end{table} 
\begin{figure*}[tb]
   \centering
   \includegraphics[width=0.45\linewidth]{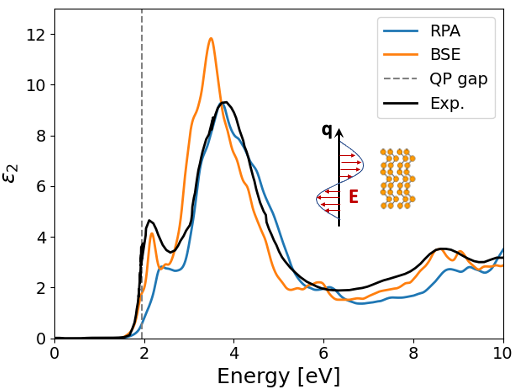}
   \includegraphics[width=0.45\linewidth]{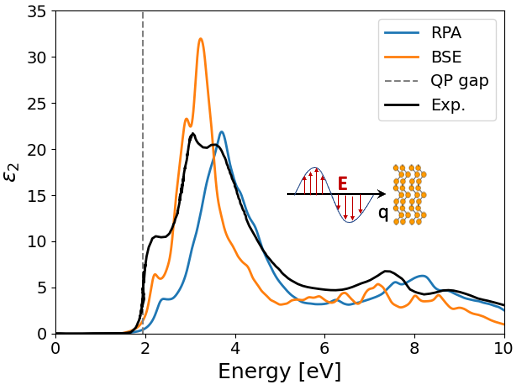}
    \caption{Optical absorption spectra calculated with BSE and RPA for electric fields polarized perpendicular (left) and parallel (right) to the selenium chains. The vertical dashed lines indicate the direct band gap. The experimental data are taken from Ref. \cite{Stuke1970}.}
    \label{fig:polarisabiliy}
\end{figure*}

Both the direct and indirect gaps play a crucial role for the performance of PV materials and it is thus of vital importance to be able to provide an accurate prediction for the band gaps. In Table \ref{tab:band_gaps} we compare the predictions of PBE, HSE06, SCAN, mBEEF and G$_0$W$_0$. From optical absorption experiments the optical gap is roughly 1.95 eV \cite{Stuke1970, Nielsen2022}. This should be compared to the direct gaps of Table \ref{tab:band_gaps} and we see that HSE06 yields a value just below the experimental gap, whereas G$_0$W$_0$ is in excellent agreement with the measured gap. The Kohn-Sham gap from PBE is severely underestimated whereas SCAN and mBEEF yields values closer to the HSE06 result. The indirect gap has been measured to be 1.80 eV \cite{ZETSCHE19691425} and 1.85 eV \cite{PhysRevLett.42.264}, which is in good agreement with our HSE06 and G$_0$W$_0$ results. The present G$_0$W$_0$ gap is also in good agreement with a previous calculation, which gave 1.74 eV \cite{PhysRevLett.114.206401}. 

We note that all calculations here were carried out using the experimental crystal structure. We have tested the effect of optimizing lattice constants and atomic positions with PBE and find the lattice parameters $a$= 4.511 {\AA} and $c$=5.048 {\AA} implying a change of 0.15 {\AA} and 0.094 {\AA} with respect to experimental lattice constants. The gaps obtained from each of the methods using the PBE optimized structures are found to vary by less than 20 meV compared to the values stated in Table \ref{tab:band_gaps}. This is in agreement with Ref. \cite{stoliaroff2021impact} where it was found that PBE yields reasonable lattice constants whereas van der Waals D3 corrections to PBE worsens the results. Similarly, SCAN was found to produce good lattice constants, which becomes worse with the inclusion of D3 corrections, and HSE06 was shown to significantly overestimate the in-plane lattice constant.

The carrier mobilities quantify how effectively excited electrons and holes are transported across PV junctions under illumination. A low mobility may thus be detrimental for PV applications even if a material exhibits near perfect optical absorption. In general, the mobility will depend on a given sample and is strongly influenced by the amount and type of impurities. Direct calculation of mobilities including scattering from phonons, impurities, and other sources is beyond the scope of the present work. However, if one assumes an effective relaxation time $\tau$ one may calculate the conductivity tensor in the Drude model as
\begin{align}\label{eq:conductance}
\sigma_{ij}=ne^2\tau \tilde{m}^{-1}_{ij}
\end{align}
where $e$ is the electron charge, $n$ is the carrier density and $\tilde{m}_{ij}$ is the effective transport mass tensor. Due to the symmetries of selenium the conductivity tensor must be diagonal and isotropic in the plane perpendicular to the Se chains. For a particular VBM or CBM the inverse mass tensor is simply the curvature tensor evaluated at the extremal $k$-point, but unless the band edge is located at a high symmetry point this mass tensor will not reflect the symmetries of the crystal. Indeed, for the case of degenerate valleys in the Brillouin zone Eq. \eqref{eq:conductance} should be replaced by a sum over valleys, each carrying a fraction of the total carrier density $n$. It is then clear that the transport mass tensor in Eq. \eqref{eq:conductance} is given by
\begin{align}
\tilde{m}^{-1}_{ij}=\frac{1}{N_v}\sum_{\alpha=1}^{N_v}{m}^{-1}_{\alpha,ij},
\end{align}
where $N_v$ is the number of degenerate valleys and ${m}^{-1}_{\alpha,ij}$ is the mass tensor of valley $\alpha$. Since the mass tensors in different valleys are related by symmetry it is sufficient to calculate the mass tensor of a single valley. In the SM \cite{SM} we show the quadratic fit of bands for a particular VBM and CBM along the eigendirections of the mass tensor. From this mass tensor we may then calculate the transport mass tensor, which exhibits the same symmetries as the conductivity. For holes (h) and electron (e) we obtain $\tilde{m}^\mathrm{h}_\perp=0.83$, $\tilde{m}^\mathrm{h}_\parallel=0.27$, $\tilde{m}^\mathrm{e}_\perp=0.38$ and $\tilde{m}^\mathrm{e}_\parallel=0.18$, where $\perp$ and $\parallel$ indicates directions perpendicular and parallel to the Se chains respectively. As expected the masses are lower along the chains due to strong dispersion in this direction. In general, it will thus be advantageous to orient PV devices based on single crystals of selenium such that carriers are transported along the Se chains. For a particular defect, the relaxation time may be calculated from first principles within the T-matrix formalism \cite{Kaasbjerg2020}, but this will not be pursued further here.

The optical absorption of bulk materials is largely governed by the imaginary part of the dynamical dielectric constant $\varepsilon_2(\omega)$. This quantity is commonly calculated in the RPA, which comprises a decent approximation for insulators if the band gap is not too large. It is, however, well known that the RPA is not able to capture excitonic effects and for bulk silicon the absorption at the band edge is strongly underestimated by the RPA \cite{PhysRevLett.80.4510}. To elucidate the excitonic effects in selenium we have calculated $\varepsilon_2(\omega)$ from the RPA as well as the BSE, which provides a good a account of excitonic effects \cite{Onida2002}. The response is evaluated at the PBE eigenvalues and eigenstates, where we have shifted the conduction bands in order to match the direct G$_0$W$_0$ gap. In Fig. \ref{fig:polarisabiliy} we show the optical absorption for electric fields polarized parallel (along $z$) and perpendicular (along $x$) to the Se chains. Comparing the RPA and BSE results we see that the results are qualitatively similar, but BSE yields an absorption edge that is shifted to 50 meV below the quasi-particle gap. In addition, for light polarized parallel to the chains, there is a significant shift of spectral weight to lower energies. As expected, one obtains a much stronger absorption for fields  polarized parallel to the chains due to the high hybridization and mobility in this direction (large dipole matrix elements). The absorption perpendicular to the chains is roughly a factor of three smaller. The spectra for both polarization directions are in good agreement with experimental results from Ref. \cite{Stuke1970}. In particular, for $z$-polarized light the experimental spectrum is dominated by a broad double peak between 3 and 4 eV, which reaches a maximum at $\varepsilon_2=22$ and the experimental absorption edge has a small shoulder at 2.2 eV, which is also found in the BSE calculations here. For $x$-polarized light the experimental absorption spectrum is composed of a distinct peak at 2.1 eV reaching a maximum of $\varepsilon_2=5$, followed by a broader peak centered at 3.9 eV with a maximum of $\varepsilon_2=10$. This main peak is in better agreement with our RPA calculations, whereas BSE yields a much better agreement with the absorption edge. We note that the absorption along the chains is roughly a factor of two smaller than the absorption coefficient in pristine silicon where $\varepsilon(\omega)$ reaches a plateau between 35 and 45 in the vicinity of the absorption edge \cite{PhysRevLett.80.4510}.

\subsection{Point Defects}

    \label{fig:gapstates_v_Se}
Point defects may or may not give rise to localized states in the band gap, which can act as recombination centers through non-radiative decay of electron-hole pairs \cite{abakumov1991nonradiative,alkauskas2016tutorial} and typically reduce the efficiency in solar energy conversion to values below the Shockley-Queisser limit \cite{shockley1961detailed}. In addition, charged defects act as scattering centers with long-range Coulomb interaction and may be detrimental for efficient separation of excited holes and electrons.

One can get a rough idea of the influence of a particular defect from the position of defect levels (Kohn-Sham eigenenergies) in the band gap. Deep levels (far from band edges) are not expected to introduce significant doping whereas shallow levels (close to the band edges) typically have a strong influence on carrier concentrations. In the SM \cite{SM} we show the density of states in different charge states for the Se vacancy (comparing mBEEF, SCAN and PBE) as well as the interstitial Se defect and extrinsic defects. In the SM \cite{SM} we also show the projected density of states, which allows one to resolve the defect levels in contributions from the extrinsic atomic orbitals and bulk Se states.

We have performed a systematic computational assessment of the thermodynamic properties of defects as well as the possibly detrimental influence on PV applications due to non-radiative recombination. We study both intrinsic defects (a vacancy and an interstitial) and extrinsic defects where Se atoms are substituted with elements from group 4, 5, 6 or 7. We start by calculating the formation energies, charge transition levels, and equilibrium carrier concentrations. Finally, we calculate carrier capture coefficients and non-radiative recombination rates for the selenium vacancy.

\subsubsection{Defect formation energies}
\begin{figure}
    \centering
    \includegraphics[width=0.9\linewidth]
    {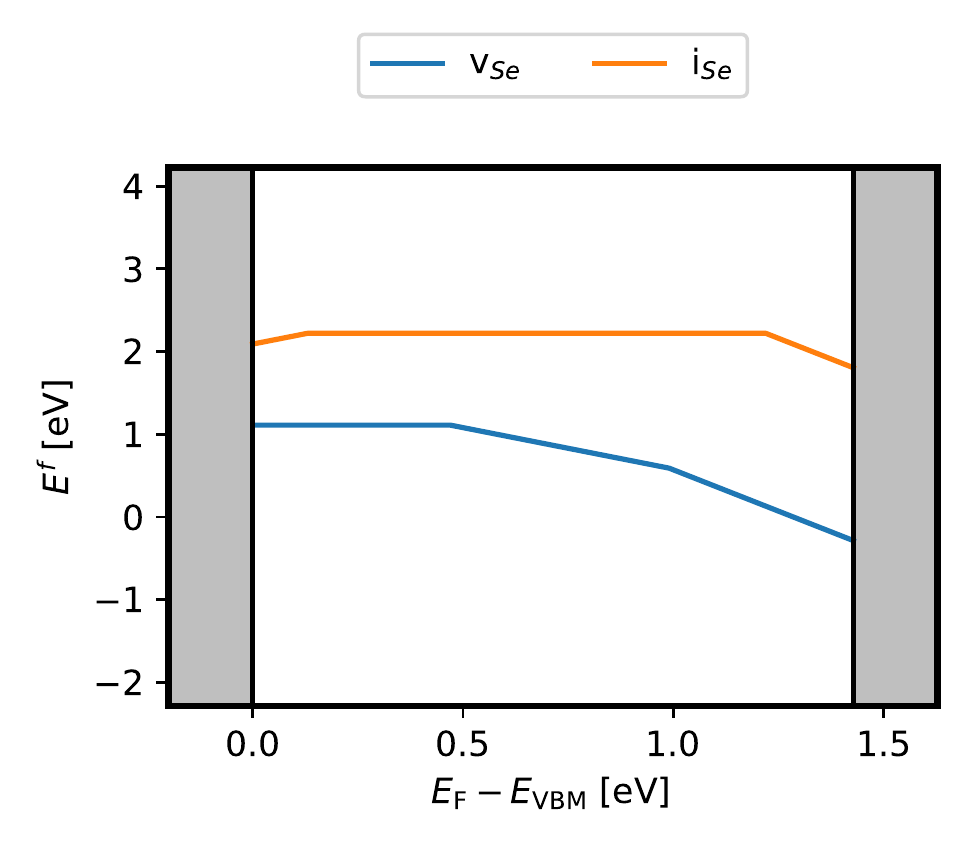}
    \caption{Formation energies of the interstitial defect (i$_\mathrm{Se}$) and the vacancy (V$_\mathrm{Se}$) obtained with mBEEF.}
    \label{fig:inter}
\end{figure}
\begin{figure*}
    \centering
    \includegraphics[width=0.45\linewidth]
    {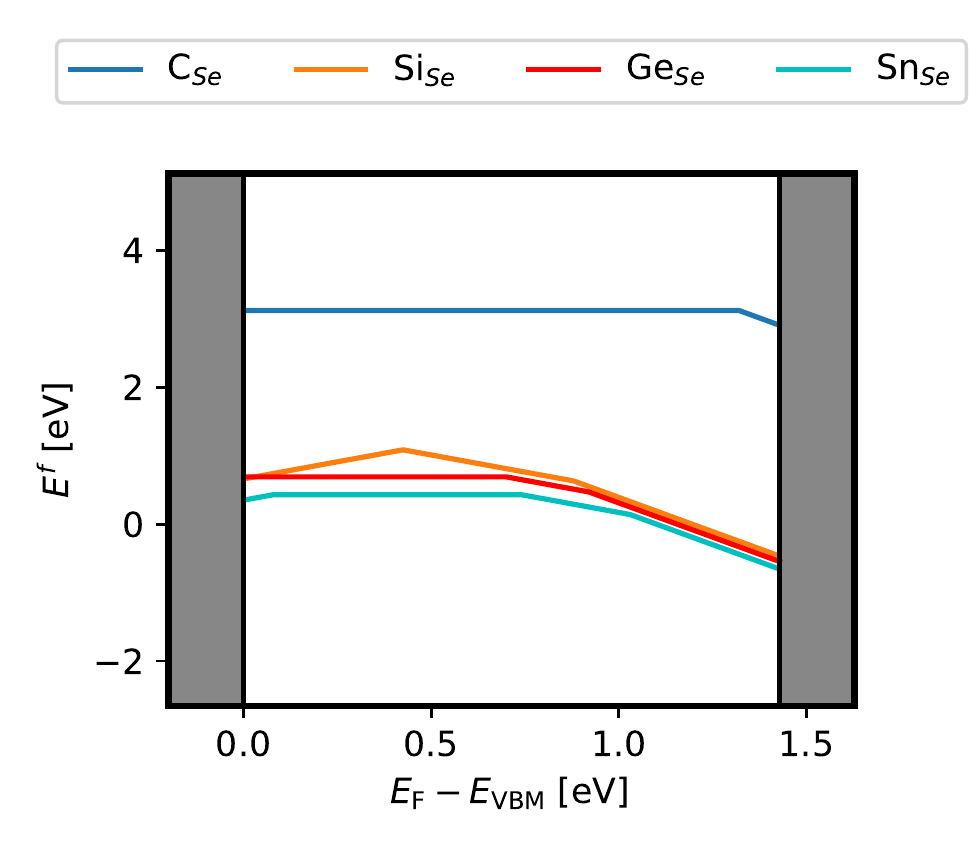}
    \includegraphics[width=0.45\linewidth]
    {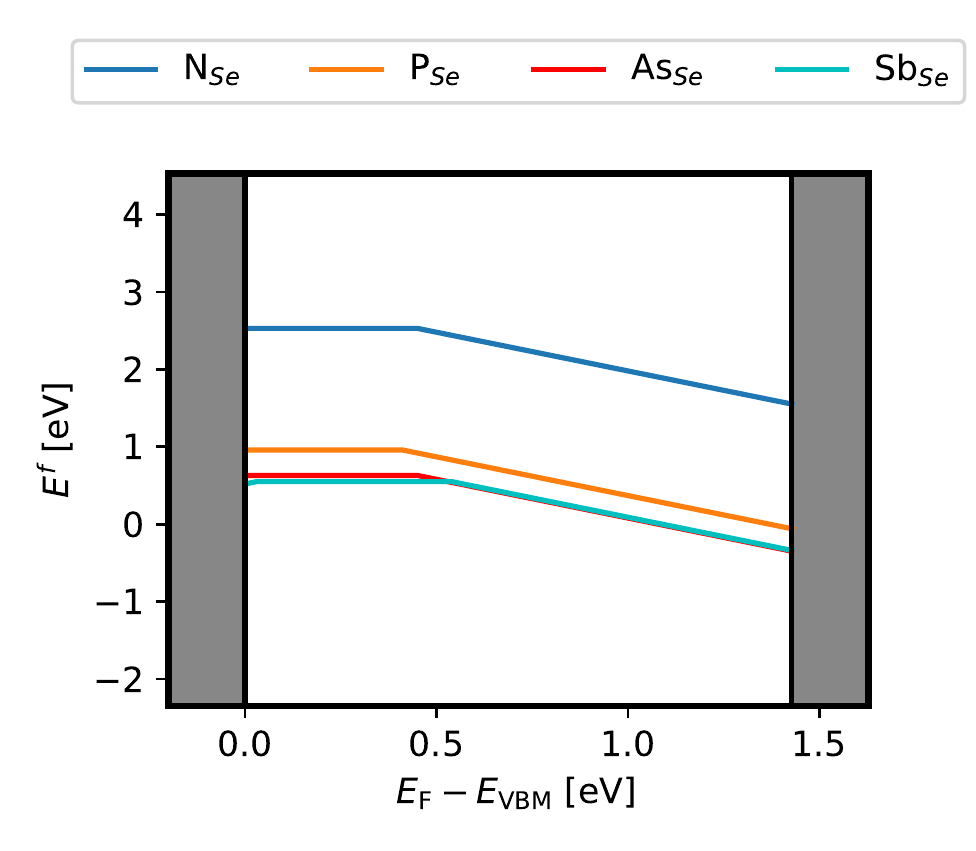}\\
    \includegraphics[width=0.45\linewidth]
    {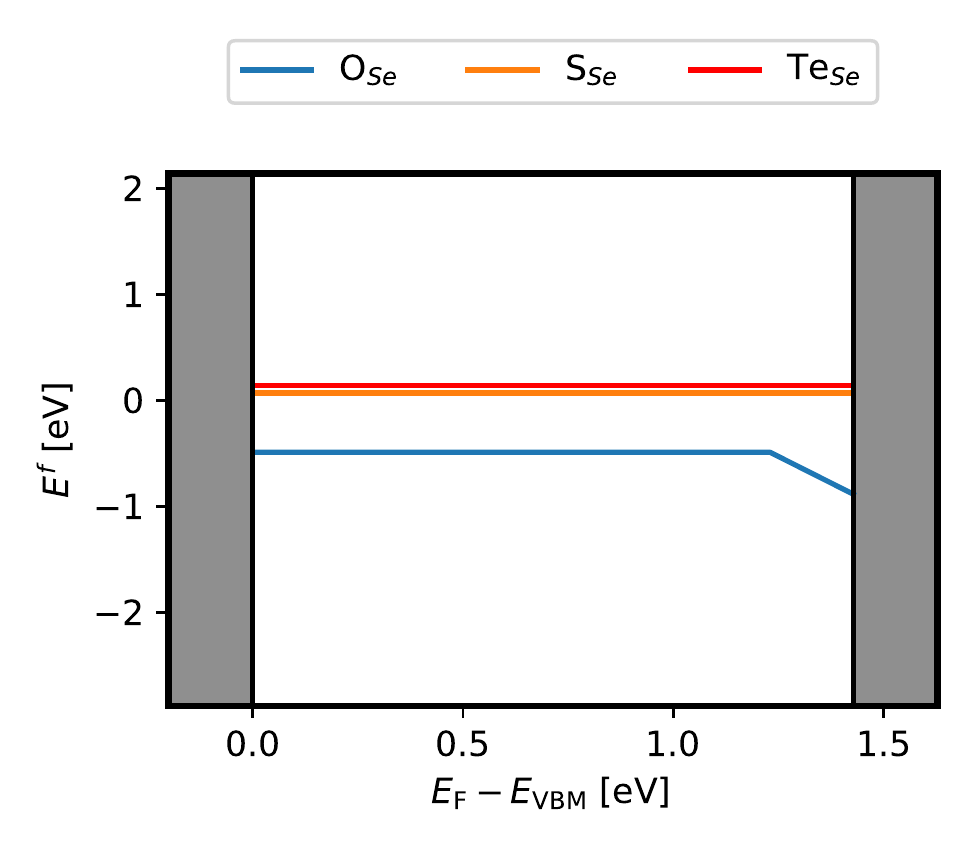}
    \includegraphics[width=0.45\linewidth]
    {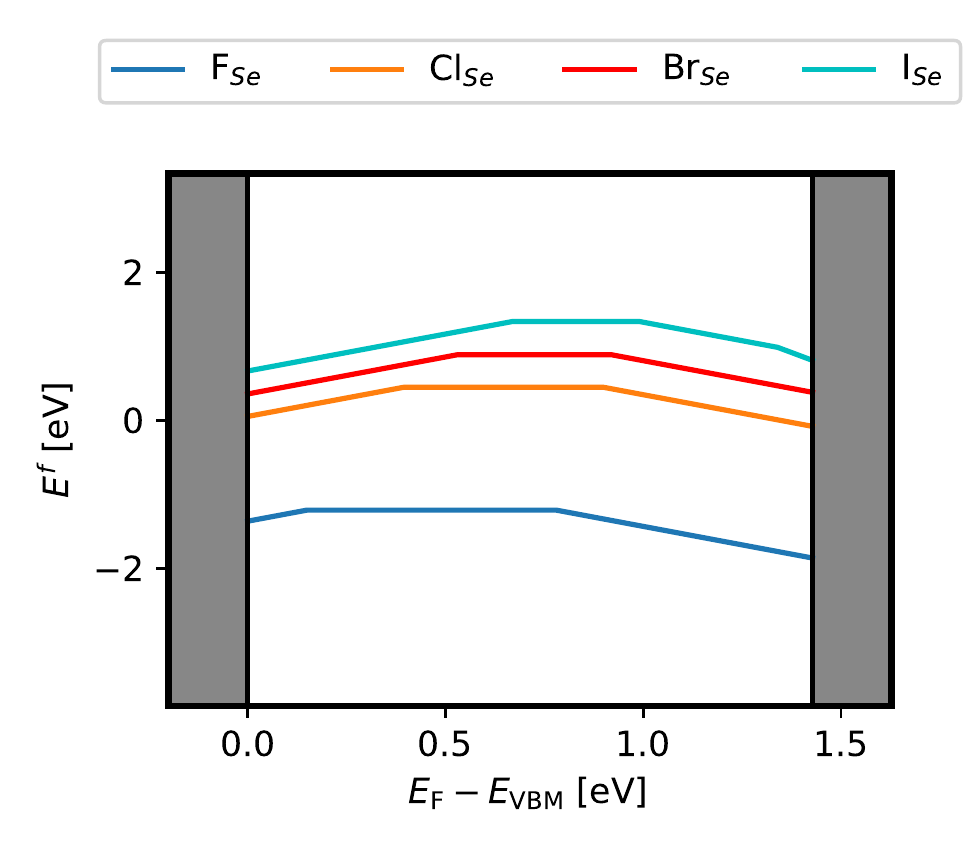}
    \caption{Formation energy diagrams of group 4, 5, 6, and 7 substitutional defects in selenium obtained with mBEEF. The Formation energies are calculated from Eq. \eqref{eq:eform} referenced to the standard states of the elements and shown as a function of the Fermi level.}
    \label{fig:formation energies}
\end{figure*}
\begin{table}[b]
\centering
    \begin{tabular}{|l|r|r|r|r|r|} 
\hline
 &$q=-2$&$q=-1$&$q=0$&$q=1$&$q=2$ \\ \hline
mBEEF&2.57&1.58&1.11&1.30&\\ \hline
SCAN&2.65&1.64&1.25&1.46&\\ \hline
PBE&2.05&1.33&0.95&1.20&\\ \hline
HSE06&3.89&2.24&1.32&0.87&0.72\\ \hline
\end{tabular}
    \caption{Formation energies of the vacancy in different charge states obtained with mBEEF, SCAN and PBE.}\label{tab:form}
\end{table}     
The thermodynamical properties of defects are determined by the formation energies in the different charge states, which may be calculated from Eq. \eqref{eq:eform}. The correction for charged defect calculations appearing in Eq. \eqref{eq:eform} relies on the subtraction of spurious electrostatic interactions between localized charges in repeated images of the super cell, which scales with the inverse dielectric constant. In principle, one should use a dielectric constant that is consistent with the functional applied for the defect formation energies, but for simplicity we have used the RPA dielectric constant calculated at PBE eigenenergies and orbitals for all calculations. We obtain $\epsilon^\perp=6.79$ and $\epsilon^\perp=11.88$ for electric fields perpendicular and parallel to the selenium chains respectively. The model used to correct for the long range interactions assumes an isotropic system and we have thus used a spherical average, which gives $\tilde\epsilon=8.49$. At the end of the day, the accuracy of the corrections needs to be assessed by converging the result with respect to super cell size and in the SM \cite{SM} we validate the accuracy of charged defect calculations in the $3\times3\times3$ super cell for the case of a Se vacancy in the $q=-2$ state.

We start by considering the intrinsic defects where a selenium atom is either removed (V$_\mathrm{Se}$) or added at an interstitial site (i$_\mathrm{Se}$). In Table \ref{tab:form} we present the formation energies of the vacancy obtained with PBE, SCAN, mBEEF and HSE06 in the different charge states (with $E_\mathrm{F}=E_\mathrm{VBM}$). The SCAN and mBEEF results are rather similar in the sense that all charge states differ in formation energy by roughly 0.1 eV. This implies that the the two functional predict similar positions of the VBM and the two formation energy curves are simply shifted by $\sim0.1$ eV. In contrast, the HSE06 results exhibit differences with respect to the meta-GGAs that are largely dependent on the charge state. The $q=0$ formation energy is similar to the SCAN result whereas the $q=1$ is somewhat smaller and the $q=-2$ is much larger. This implies that the VBM is predicted to be much lower (with respect to charge transition levels) compared to the meta-GGAs. Such differences may lead to a strong dependence on the applied functional when considering charge balance and non-radiative recombination rates. Below, we will focus on the mBEEF results for formation energies and recombination rates. In the SM \cite{SM} we perform a full calculation of the recombination rates induced by Se vacancies using the HSE06 and show that it does not change the recombination rates despite the large differences in formation energies.

The formation energies of the intrinsic defects are shown in Fig. \ref{fig:inter} as a function of Fermi energy, where we only show the charge state of lowest energy. It is clear that the vacancy generally has a much lower energy than the interstitial and is thus expected to dominate the distribution of intrinsic defects. Since the positive charged states do not acquire the lowest energy anywhere in the gap, the vacancy will in general act as acceptor for electrons and induce $p$-doping in the material. As will be shown below, the vacancy also tends to govern the charge balance and carrier densities when extrinsic defects are included. This is in agreement with previous studies, which confirm the intrinsic and extrinsic \textit{p}-type behaviour of selenium in both amorphous \cite{kolobov1996origin} and crystalline \cite{stoliaroff2021impact,abdullaev1965effect} forms.

In the case of extrinsic defects, the formation energies depend on the choice of reference state (chemical potentials) of the involved elements. To start with, we will take the atomic chemical potentials as those obtained from the standard states. Other chemical environments are, however, possible and are likely to be more relevant for realistic growth conditions. Such change in reference state will introduce a constant shift of the formation energy curve and will be discussed below when considering the charge balance. The formation energies of substitutional defects from groups 4, 5, 6, and 7 are shown in Fig. \ref{fig:formation energies}, where we again only include the charge state of lowest energy at a given Fermi energy. The charge transition levels roughly correspond to the defect levels appearing in the gap in the DOS  for the individual defects (see SM \cite{SM} figures 5-6). The Te and O defects, for example, only have states in the gap for the $q=-2$ and $q=-1$ states respectively and only the O defect exhibits a charge transition level (0/-1) close to the CBM. In contrast, the case of Cl has states in the gap for $q=-1,0,1$ charge states and shows 1/0 and 0/-1 charge transition levels in the gap. The formation energies are in qualitative (but not quantitative) agreement with those of Ref. \cite{stoliaroff2021impact} where the Sb$_\mathrm{Se}$ and Br$_\mathrm{Se}$ were studied with the SCAN functional. For the case of Br$_\mathrm{Se}$ we find the 1/0 transition at $\sim0.5$ eV above the VBM and the 0/-1 transition at $\sim0.9$ eV whereas Ref. \cite{stoliaroff2021impact} find the 1/0 transition to lie $\sim0.3$ eV, but no 0/-1 transition in the gap. For the Sb$_\mathrm{Se}$ defect we find the 0/-1 transition positioned at $\sim0.5$ eV and  Ref. \cite{stoliaroff2021impact} find it to be located at $\sim0.1$ eV above the VBM. While the formation energy curves are in qualitative agreement there are certainly quantitative differences. In the SM \cite{SM} Tables 1 and 2 we summarize all formation energies calculated with SCAN and mBEEF and the results are seen to be rather similar. The differences between the present results and those of Ref. \cite{stoliaroff2021impact} thus cannot originate from the choice of functionals, but is likely related to a correction scheme for the VBM relative to the charge transition levels applied in Ref. \cite{stoliaroff2021impact}.

With the exception of C$_\mathrm{Se}$ and N$_\mathrm{Se}$ none of the formation energies exceed $\sim 1$ eV and most of the extrinsic defects would be expected to be present if one considers thermodynamic equilibrium with the standard reference states. There is of course significant kinetic barriers for generating defects and the concentration of extrinsic defects in a given sample will depend crucially on the presence of elements during synthesis. Several of the substitutional defects have negative formation energies for acceptor levels ($q<0$) when the Fermi level approach the CBM, but since the carrier concentration is likely to be dominated by intrinsic defects 
the Fermi level becomes pinned close to the VBM and the negative charge states will be less populated than the neutral and positive states for all defects. 

The cases of F\textsubscript{Se} and O\textsubscript{Se} have negative formation energies for all charge states and are thus predicted to destabilize the bulk phase. However, the formation energies depend crucially on the chemical potentials chosen for the substitutional elements and the results of Fig. \ref{fig:formation energies} are specific to the choice of standard reference states. For a proper thermodynamic analysis one needs to take into account the possibility of forming secondary phases, and the chemical potentials may acquire shifts that are bounded by the formation energies (per defect element) of secondary phases \cite{freysoldt2014first}. We refer to the SM \cite{SM} for details on this and a table of secondary phase formation energies. In Table \ref{tab:form_energies_shift} we state the calculated formation energy shifts, which must be added to the formation energies (calculated with respect to standard references) when considering Se rich conditions, which are likely to provide a more realistic description of synthesis conditions. The largest shifts are those involving F and O, which originate from the high stability of secondary phases. It can be seen that for both F\textsubscript{Se} and O\textsubscript{Se} the formation energies become positive when considering the Se rich limit.

\subsubsection{Charge balance and carrier concentrations}
Given the formation energies of defects in various charge states the Fermi level may be determined by the charge neutrality condition \cite{buckeridge2019equilibrium,bertoldo2022quantum}
\begin{equation}\label{eq:charge_neutrality}
\sum_\mathrm{X}\sum_q qC[\mathrm{X}^q, E^f, T] = n_0 - p_0
\end{equation}
where $C[\mathrm{X}^q, E^f, T]$ is the concentration of a given defect X in charge state $q$, and $n_0$ ($p_0$) is the electron (hole) carrier density. The defect concentration $C[\mathrm{X}^q, E^f, T]$ is determined by the Boltzmann distribution, which depends on the formation energy as well as the temperature. Both the formation energies and the carrier densities depend on the Fermi level and this equation must thus be solved self-consistently. Since the formation energies also depend on the choice of reference for atomic chemical potentials one may solve for the self-consistent Fermi level and the carrier densities for different chemical environments. We refer to the SM \cite{SM} for details on the charge balance equation \eqref{eq:charge_neutrality}.

In the case of intrinsic defects only, the solution of the charge balance equation at T=300 K yields $E_\mathrm{F}-E_\mathrm{VBM}=0.72$ eV and we obtain carrier densities of $p_0=1.4\times10^8$ cm$^{-3}$ and $n_0=5.2\times10^6$ cm$^{-3}$. While the Fermi level is positioned very close to the middle of the gap the density of holes is more than an order of magnitude larger than the electron density. This is due to the much larger density of states at the VBM compared to the CBM, which is expected from the effective mass calculations presented above. The net positive charge of the carriers is fully compensated by the vacancies in the $q=-1$ charge state, which has roughly the same density as the holes. Other charge states of the vacancy as well as the interstitial defects have densities that are at least four orders of magnitude lower. 

However, the defects are likely to form during growth of the material, which happens at much larger temperatures and if one assumes that the defects are frozen in during synthesis, one should solve the charge balance equation at room temperature with a fixed density of defects corresponding to thermodynamic equilibrium at the growth conditions. The growth temperature of trigonal selenium can be taken to be 500 K \cite{todorov2017ultrathin}, which gives a density of vacancies of $2.8\times10^{13}$ cm$^{-3}$. With this density the charge balance at T=300 K yields a Fermi energy of 0.44 eV above the VBM and a hole doping of $p_0=6.8\times10^{12}$ cm$^{-3}$. This number is exactly balanced by the number of vacancies in charge state $q=-1$, and the majority of the vacancies thus reside in the neutral state. The density of electrons in the conduction band also acquires a much lower value of $n_0=1.0\times10^2$ cm$^{-3}$.

The inclusion of extrinsic defects does not change the charge balance much. The vacancy still plays a dominating role in the charge balance. We have tested the inclusion of substitutional defects involving O, Cl, F under selenium-rich conditions (see SM \cite{SM}). The density of Cl defects is on the same order of magnitude as vacancies, but the hole-doping and Fermi level does not change much. Nevertheless, such extrinsic defects may play a crucial role in the non-radiative recombination to which we will now turn.
\begin{table}[tb]
    \centering
    \begin{tabular}{ | c | r | r | r | r}
    \hline
    Defect species & mBEEF & SCAN  & PBE\\ \hline
    P\textsubscript{Se} & 0.20 & 0.16 & 0.167 \\
    As\textsubscript{Se} & 0.35&  0.34 & 0.29\\
    Sb\textsubscript{Se} & 0.55&  0.57 & 0.66\\
    O\textsubscript{Se} & 1.16 & 1.13 & 1.05 \\
    F\textsubscript{Se}& 2.14  & 2.21 & 2.23 \\
    Cl\textsubscript{Se}& 0.32 & 0.29 & 0.45 \\
    \hline
    \end{tabular}
        \caption{Defect formation energy shifts (in eV per defect atom) calculated with mBEEF for Se-rich conditions.} 
    \label{tab:form_energies_shift}
\end{table}

\subsection{Nonradiative carrier capture rates and solar cell efficiency}
\begin{figure*}
\centering
\includegraphics[width=0.8\linewidth]{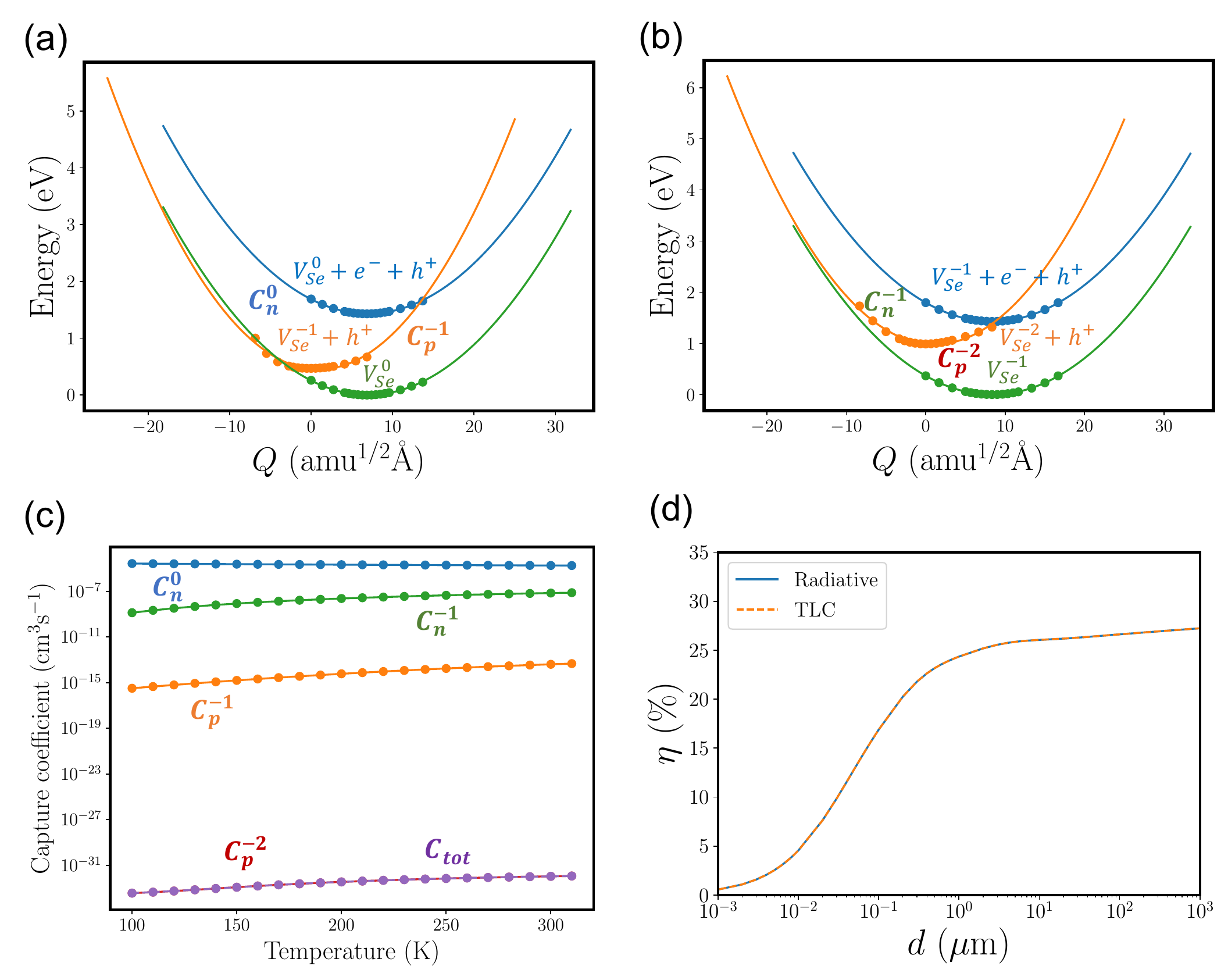}
\caption{Configuration coordinate diagrams for (a) (0/-1) and (b) (-1/-2) charge transitions of V$_\mathrm{Se}$. (c) Non-radiative capture coefficients for V$_\mathrm{Se}$ and (d) associated trap limited conversion efficiency (TLC) vs. film thickness ($d$) plot for thin-film Se solar cell. The radiative efficiency is also plotted for comparison, but falls identically on the line associated with the TLC.}
\label{fig:nonradiativePV}
\end{figure*}
The main material parameter used to quantify the potential of a photovoltaic material in the radiative limit is its absorption coefficient (see related discussion in SM \cite{SM}. For Se, we calculate the direct absorption (vertical transitions) using RPA and the phonon-assisted absorption is calculated using second order perturbation theory as explained in Ref. \cite{kangsabanik2022indirect}. After putting all the relevant quantities into Eq.~S12 of the SM \cite{SM} we obtain the thickness dependent radiative efficiency ($\eta_\mathrm{rad}$). In the thin-film limit ($\approx$ 1$\mu$m) we obtain a radiative efficiency of $\approx 25\%$ (see Fig. \ref{fig:nonradiativePV}(d)). However, this value is significantly higher than the experimentally measured efficiency of 6.5\% \cite{todorov2017ultrathin,Nielsen2022}. This indicates that, there might be significant non-radiative recombination taking place inside the material governed by the deep level defects formed during synthesis. 

The presence of deep level defects can lead to non-radiative Shockley-Read-Hall (SRH) recombination \cite{shockley1952statistics, hall1952electron}, which is known to limit the photovoltaic efficiency of semiconductors such as CdTe \cite{kavanagh2021rapid} and Kesterite \cite{kim2020upper}. Here we explore how non-radiative SRH recombination via intrinsic defects influence the power conversion efficiency in Se. In case of elemental Se, the two possible intrinsic defects are V$_\mathrm{Se}$ and i$_\mathrm{Se}$. However, the high formation energy of i$_\mathrm{Se}$ (see Fig. \ref{fig:inter}) renders this defect unlikely to be present in any significant concentration, and will therefore not be considered further here. In contrast, the V$_\mathrm{Se}$ defect was found to be present in large amounts and it has two charge state transitions deep inside the band gap, both of which will contribute to non-radiative SRH recombination. 

In Fig. \ref{fig:nonradiativePV}(a) and \ref{fig:nonradiativePV}(b) we show the configuration coordinate diagram of carrier capture at two different charge transitions  ((0/-1) and (-1/-2), respectively) for the V$_\mathrm{Se}$ defect. More details about the related methodology can be found in the SM \cite{SM}. For the first case (Fig. \ref{fig:nonradiativePV}(a)) it can be assumed that we have a neutral defect V$_\mathrm{Se}^0$ with an electron in the conduction band and a hole in the valence band. At first, V$_\mathrm{Se}^0$ captures an electron and relaxes to the V$_\mathrm{Se}^{-1}$ state. V$_\mathrm{Se}^{-1}$ then captures a hole and consequently relaxes back to the neutral charge state, completing the recombination process. Semi-classically the energy barriers, determined by the intersection of the potential energy surfaces of the initial and the final state, can provide a first estimate of the rate of carrier capture. However, for quantitative estimates one must employ a full quantum mechanical treatment e.g. using the formalism of Ref. \cite{alkauskas2014first} as described in SM \cite{SM}. Now, for a defect such as V$_\mathrm{Se}$ showing multiple charge state transitions inside the band gap, one can calculate the total capture rate ($C_\mathrm{tot}$) by solving the detailed balance of carrier capture under steady-state conditions \cite{sah1958electron, alkauskas2016role}. In this case $C_\mathrm{tot}$ can be expressed as,
\begin{equation}
	C_\mathrm{tot}=\frac{C_n^{-1}+C_p^{-1}}{1+\frac{C_n^{-1}}{C_p^{-2}}+\frac{C_p^{-1}}{C_n^0}}.
	\label{eq:CCm}
	\end{equation}
Here $C_n^0$ denotes an electron ($n$) capture from an initial neutral defect. Similarly the hole ($p$) capture by a negatively charged defect is expressed as $C_p^{-1}$. The calculated non-radiative capture coefficients are shown in Fig. \ref{fig:nonradiativePV}(c). Although, $C_n^{-1}$, $C_p^{-1}$, and $C_n^0$ show moderate to high values, $C_p^{-2}$ is extremely small and dominates the expression of $C_\mathrm{tot}$ as can be seen from Eq. \eqref{eq:CCm}. $C_p^{-2}$ thus constitutes the major rate limiting step and practically coincides with the value of $C_\mathrm{tot}$. Intuitively, this can be explained by a very high energy barrier for hole capture by the $q=-2$ charge state of V$_\mathrm{Se}$, for which  the two potential energy surfaces do not even intersect. Combining this with the defect concentration of V$_\mathrm{Se}$ we may calculate the SRH recombination coefficient and thus the trap limited conversion efficiency (TLC) \cite{kim2020upper}. This is shown in Fig. \ref{fig:nonradiativePV}(d) as a function of film thickness and it is observed that the efficiency saturates when the thickness is 0.1-1 $\mu$m. The saturation itself is in good agreement with the measurements in Ref. \cite{youngman2021semitransparent} where saturation was observed at 0.3 $\mu$m. As expected from the low carrier capture rate, we see that the TLC is exactly the same as the radiative efficiency. On this basis we conclude that the intrinsic defects cannot explain the reduced efficiency observed experimentally. One might worry if this result is an artifact of the mBEEF formation energies, which could be too inaccurate for such calculations. In general, it is not {\it a priori} clear how reliable the predictions of DFT are with respect to formation energies since these depend on the position of the VBM. Since the HSE06 tend to predict Kohn-Sham band gaps in good agreement with experiments it is generally believed to be superior for defect calculations. We have thus repeated the calculation of formation energies and recombination rates with HSE06 (shown in the SM \cite{SM}) and find the same conclusion - the intrinsic defects do not give rise to non-radiative recombination.

The next natural step would thus be to calculate the contribution to SRH recombination from extrinsic defects, but in order to obtain accurate predictions this requires detailed knowledge of the extrinsic defect concentrations, which are very sensitive to the chemical environment during synthesis. Finally, we have not considered the possibility of defect complexes beyond atomic substitutions. Such complexes could play a role in recombination processes, but a systematic study of associated formation energies and recombination rates is beyond the scope of this work.

\section{Conclusion and Outlook}\label{sec:conclusion}
We have presented first-principles calculations of the electronic properties of pristine trigonal selenium and investigated the role of defects with respect to carrier concentrations and non-radiative recombination. By applying GW and BSE we obtain excellent agreement with the experimental values for band gap and optical absorption respectively. In addition, effective mass calculations reveal that pristine selenium exhibits mobilities that are higher or comparable to those of silicon. 

We then investigated the role of intrinsic defects, which were shown to be dominated by vacancies, that act as acceptors and give rise to $p$-doping. The calculated doping levels originating from vacancies are, however, orders of magnitude lower than those observed experimentally \cite{Nielsen2022}. In addition, non-radiative recombination rates from vacancies are vanishingly small and cannot explain the severe reduction of photovoltaic efficiency observed in selenium-based photovoltaic devices \cite{nakada1985polycrystalline, todorov2017ultrathin, Nielsen2022}.

These observations imply that the presence of extrinsic defects is likely to be decisive for the efficiency and we have analyzed the properties of substitutional defects involving elements from groups 4, 5, 6 and 7. In that case we find that the doping levels do not change much, but several of the defects yield deep levels in the gap that may lead to significant non-radiative recombination. This could explain the severe reduction of efficiency in selenium-based single junction devices and implies that the efficiency can be increased significantly by applying synthesis conditions that minimize the probability of introducing extrinsic defects.

\section{Acknowledgements}
J.K and K.S.T acknowledge funding from the European Research Council (ERC) under the European Union’s Horizon 2020 research and innovation program Grant No. 773122 (LIMA) and Grant agreement No. 951786 (NOMAD CoE). K. S. T. is a Villum Investigator supported by Villum Fonden (grant no. 37789). The Villum Center for Science of Sustainable Fuels and Chemicals, which is supported by the Villum Fonden research Grant No. 9455, is acknowledged by K.W.J. and H.M. 

\bibliographystyle{unsrt}
\bibliography{references}

\begin{thebibliography}{10}

\bibitem{smith1873action}
Willoughby Smith.
\newblock The action of light on selenium.
\newblock {\em Journal of the Society of Telegraph Engineers}, 2(4):31--33,
  1873.

\bibitem{ferry2020challenges}
DK~Ferry, SM~Goodnick, VR~Whiteside, and IR~Sellers.
\newblock Challenges, myths, and opportunities in hot carrier solar cells.
\newblock {\em Journal of Applied Physics}, 128(22):220903, 2020.

\bibitem{nakada1985polycrystalline}
Tokio Nakada and Akio Kunioka.
\newblock Polycrystalline thin-film \ch{TiO2}/\ch{Se} solar cells.
\newblock {\em Japanese journal of applied physics}, 24(7A):L536, 1985.

\bibitem{madelung2004semiconductors}
Otfried Madelung.
\newblock {\em Semiconductors: data handbook}.
\newblock Springer Science \& Business Media, 2004.

\bibitem{todorov2017ultrathin}
Teodor~K Todorov, Saurabh Singh, Douglas~M Bishop, Oki Gunawan, Yun~Seog Lee,
  Talia~S Gershon, Kevin~W Brew, Priscilla~D Antunez, and Richard Haight.
\newblock Ultrathin high band gap solar cells with improved efficiencies from
  the world’s oldest photovoltaic material.
\newblock {\em Nature communications}, 8(1):1--8, 2017.

\bibitem{john2017electronic}
JD~John, I~Saito, R~Toyama, J~Ochiai, T~Yamada, T~Masuzawa, DHC Chua, and
  K~Okano.
\newblock Electronic properties and potential applications of the
  heterojunction between silicon and multi-nanolayer amorphous selenium.
\newblock {\em Electronics Letters}, 53(18):1270--1272, 2017.

\bibitem{bishop2017record}
Douglas~M Bishop, Teodor Todorov, Yun~Seog Lee, Oki Gunawan, and Richard
  Haight.
\newblock Record efficiencies for selenium photovoltaics and application to
  indoor solar cells.
\newblock In {\em 2017 IEEE 44th Photovoltaic Specialist Conference (PVSC)},
  pages 1441--1444. IEEE, 2017.

\bibitem{shockley1961detailed}
William Shockley and Hans~J Queisser.
\newblock Detailed balance limit of efficiency of \textit{p-n} junction solar
  cells.
\newblock {\em J. Appl. Phys}, 32(3):510--519, 1961.

\bibitem{polman2016photovoltaic}
Albert Polman, Mark Knight, Erik~C Garnett, Bruno Ehrler, and Wim~C Sinke.
\newblock Photovoltaic materials: Present efficiencies and future challenges.
\newblock {\em Science}, 352(6283):aad4424, 2016.

\bibitem{youngman2021semitransparent}
Tomas~H Youngman, Rasmus Nielsen, Andrea Crovetto, Brian Seger, Ole Hansen,
  Ib~Chorkendorff, and Peter~CK Vesborg.
\newblock Semitransparent selenium solar cells as a top cell for tandem
  photovoltaics.
\newblock {\em Solar RRL}, 5(7):2100111, 2021.

\bibitem{Nielsen2022}
Rasmus Nielsen, Tomas~H Youngman, Hadeel Moustafa, Sergiu Levcenco, Hannes
  Hempel, Andrea Crovetto, Thomas Olsen, Ole Hansen, Ib~Chorkendorff, Thomas
  Unold, and Peter C~K Vesborg.
\newblock {Origin of photovoltaic losses in selenium solar cells with
  open-circuit voltages approaching 1 V}.
\newblock {\em J. Mater. Chem. A}, 10(45):24199--24207, 2022.

\bibitem{enkovaara2010electronic}
Jussi Enkovaara, Carsten Rostgaard, J~J{\o}rgen Mortensen, Jingzhe Chen,
  M~Du{\l}ak, Lara Ferrighi, Jeppe Gavnholt, Christian Glinsvad, V~Haikola,
  HA~Hansen, et~al.
\newblock Electronic structure calculations with gpaw: a real-space
  implementation of the projector augmented-wave method.
\newblock {\em Journal of physics: Condensed matter}, 22(25):253202, 2010.

\bibitem{blochl}
P~E Bl{\"{o}}chl.
\newblock {Projector augmented-wave method}.
\newblock {\em Phys. Rev. B}, 50:17953, 1994.

\bibitem{kresse1999ultrasoft}
Georg Kresse and Daniel Joubert.
\newblock From ultrasoft pseudopotentials to the projector augmented-wave
  method.
\newblock {\em Physical review b}, 59(3):1758, 1999.

\bibitem{Larsen2017}
Ask {Hjorth Larsen}, Jens {J{\o}rgen Mortensen}, Jakob Blomqvist, Ivano~E
  Castelli, Rune Christensen, Marcin Du{\l}ak, Jesper Friis, Michael~N Groves,
  Bj{\o}rk Hammer, Cory Hargus, Eric~D Hermes, Paul~C Jennings, Peter {Bjerre
  Jensen}, James Kermode, John~R Kitchin, Esben {Leonhard Kolsbjerg}, Joseph
  Kubal, Kristen Kaasbjerg, Steen Lysgaard, J{\'{o}}n {Bergmann Maronsson},
  Tristan Maxson, Thomas Olsen, Lars Pastewka, Andrew Peterson, Carsten
  Rostgaard, Jakob Schi{\o}tz, Ole Sch{\"{u}}tt, Mikkel Strange, Kristian~S
  Thygesen, Tejs Vegge, Lasse Vilhelmsen, Michael Walter, Zhenhua Zeng, and
  Karsten~Wedel Jacobsen.
\newblock {The atomic simulation environment—a Python library for working
  with atoms}.
\newblock {\em J. Phys. Condens. Matter}, 29(27):273002, jul 2017.

\bibitem{Gjerding.2021}
Morten Gjerding, Thorbjørn Skovhus, Asbjørn Rasmussen, Fabian Bertoldo,
  Ask~Hjorth Larsen, Jens~Jørgen Mortensen, and Kristian~Sommer Thygesen.
\newblock {Atomic Simulation Recipes -- a Python framework and library for
  automated workflows}.
\newblock {\em computational Materials Science}, 199:110731, 2021.

\bibitem{Perdew:1996ug}
JP~Perdew, K~Burke, and M~Ernzerhof.
\newblock {Generalized gradient approximation made simple}.
\newblock {\em Physical Review Letters}, 77(18):3865--3868, January 1996.

\bibitem{sun2015strongly}
Jianwei Sun, Adrienn Ruzsinszky, and John Perdew.
\newblock Strongly constrained and appropriately normed (scan) meta-generalized
  gradient approximation for exchange and correlation.
\newblock In {\em APS March Meeting Abstracts}, volume 2015, pages B24--013,
  2015.

\bibitem{wellendorff2014mbeef}
Jess Wellendorff, Keld~T Lundgaard, Karsten~W Jacobsen, and Thomas Bligaard.
\newblock mbeef: An accurate semi-local bayesian error estimation density
  functional.
\newblock {\em The Journal of chemical physics}, 140(14):144107, 2014.

\bibitem{Olsen2016a}
Thomas Olsen.
\newblock {Designing in-plane heterostructures of quantum spin Hall insulators
  from first principles: 1T'-MoS$_2$ with adsorbates}.
\newblock {\em Phys. Rev. B}, 94(23):235106, 2016.

\bibitem{heyd2003hybrid}
Jochen Heyd, Gustavo~E Scuseria, and Matthias Ernzerhof.
\newblock Hybrid functionals based on a screened coulomb potential.
\newblock {\em The Journal of chemical physics}, 118(18):8207--8215, 2003.

\bibitem{huser2013quasiparticle}
Falco H{\"u}ser, Thomas Olsen, and Kristian~S Thygesen.
\newblock Quasiparticle gw calculations for solids, molecules, and
  two-dimensional materials.
\newblock {\em Physical Review B}, 87(23):235132, 2013.

\bibitem{Onida2002}
Giovanni Onida, Lucia Reining, and Angel Rubio.
\newblock {Electronic excitations: density-functional versus many-body
  Green's-function approaches}.
\newblock {\em Rev. Mod. Phys.}, 74(2):601--659, 2002.

\bibitem{SM}
See supplemental material at [url] for details on effective mass calculations,
  defect chemical potentials, self-consistent fermi level evaluation,
  convergence of formation energies with respect to super cell size, summary of
  density of states for various defect types and details on the calculations of
  nonradiative carrier capture rates.

\bibitem{PhysRevB.104.064105}
Tom\'a\ifmmode \check{s}\else~\v{s}\fi{} Rauch, Francisco Munoz, Miguel A.~L.
  Marques, and Silvana Botti.
\newblock Defect levels from scan and mbj meta-gga exchange-correlation
  potentials.
\newblock {\em Phys. Rev. B}, 104:064105, 2021.

\bibitem{passler1976relationships}
R~P{\"a}ssler.
\newblock Relationships between the nonradiative multiphonon carrier-capture
  properties of deep charged and neutral centres in semiconductors.
\newblock {\em physica status solidi (b)}, 78(2):625--635, 1976.

\bibitem{tiedje1984limiting}
TOM Tiedje, ELI Yablonovitch, George~D Cody, and Bonnie~G Brooks.
\newblock Limiting efficiency of silicon solar cells.
\newblock {\em IEEE Transactions on electron devices}, 31(5):711--716, 1984.

\bibitem{freysoldt2011electrostatic}
Christoph Freysoldt, J{\"o}rg Neugebauer, and Chris~G Van~de Walle.
\newblock Electrostatic interactions between charged defects in supercells.
\newblock {\em Phys. Status Solidi B}, 248(5):1067--1076, 2011.

\bibitem{cherin1972refinement}
P~Cherin and P~Unger.
\newblock Refinement of the crystal structure of $\alpha$-monoclinic se.
\newblock {\em Acta Crystallographica Section B: Structural Crystallography and
  Crystal Chemistry}, 28(1):313--317, 1972.

\bibitem{marsh1953crystal}
RE~Marsh, L~Pauling, and JD~McCullough.
\newblock The crystal structure of $\beta$ selenium.
\newblock {\em Acta Crystallographica}, 6(1):71--75, 1953.

\bibitem{foss1980crystal}
Olav Foss and Vitalijus Janickis.
\newblock Crystal structure of $\gamma$-monoclinic selenium.
\newblock {\em Journal of the Chemical Society, Dalton Transactions},
  (4):624--627, 1980.

\bibitem{keller1977effect}
R~Keller, WB~Holzapfel, and Heinz Schulz.
\newblock Effect of pressure on the atom positions in se and te.
\newblock {\em Physical Review B}, 16(10):4404, 1977.

\bibitem{cherin1967crystal}
Paul Cherin and Phyllis Unger.
\newblock The crystal structure of trigonal selenium.
\newblock {\em Inorganic Chemistry}, 6(8):1589--1591, 1967.

\bibitem{PhysRevLett.114.206401}
Motoaki Hirayama, Ryo Okugawa, Shoji Ishibashi, Shuichi Murakami, and Takashi
  Miyake.
\newblock Weyl node and spin texture in trigonal tellurium and selenium.
\newblock {\em Phys. Rev. Lett.}, 114:206401, May 2015.

\bibitem{Tutihasi1967}
Simpei Tutihasi and Inan Chen.
\newblock Optical properties and band structure of trigonal selenium.
\newblock {\em Phys. Rev.}, 158:623--630, Jun 1967.

\bibitem{Stuke1970}
J.~Stuke.
\newblock {Review of optical and electrical properties of amorphous
  semiconductors}.
\newblock {\em J. Non. Cryst. Solids}, 4:1--26, apr 1970.

\bibitem{ZETSCHE19691425}
H.~Zetsche and R.~Fischer.
\newblock Photoluminescence of trigonal selenium single crystals.
\newblock {\em Journal of Physics and Chemistry of Solids}, 30(6):1425--1428,
  1969.

\bibitem{PhysRevLett.42.264}
B.~Moreth.
\newblock Two types of indirect-exciton ground states in trigonal selenium.
\newblock {\em Phys. Rev. Lett.}, 42:264--267, Jan 1979.

\bibitem{hirayama2015weyl}
Motoaki Hirayama, Ryo Okugawa, Shoji Ishibashi, Shuichi Murakami, and Takashi
  Miyake.
\newblock Weyl node and spin texture in trigonal tellurium and selenium.
\newblock {\em Physical review letters}, 114(20):206401, 2015.

\bibitem{stoliaroff2021impact}
Adrien Stoliaroff, Camille Latouche, and St{\'e}phane Jobic.
\newblock Impact of point defects on the electrical properties of selenium: A
  density functional theory investigation with discussion of the entropic term.
\newblock {\em Phys. Rev. B}, 103(9):094111, 2021.

\bibitem{Kaasbjerg2020}
Kristen Kaasbjerg.
\newblock Atomistic $t$-matrix theory of disordered two-dimensional materials:
  Bound states, spectral properties, quasiparticle scattering, and transport.
\newblock {\em Phys. Rev. B}, 101:045433, Jan 2020.

\bibitem{PhysRevLett.80.4510}
Stefan Albrecht, Lucia Reining, Rodolfo Del~Sole, and Giovanni Onida.
\newblock Ab initio calculation of excitonic effects in the optical spectra of
  semiconductors.
\newblock {\em Phys. Rev. Lett.}, 80:4510--4513, May 1998.

\bibitem{abakumov1991nonradiative}
VN~Abakumov, Vladimir~Idelevich Perel, and IN~Yassievich.
\newblock {\em Nonradiative recombination in semiconductors}.
\newblock Elsevier, 1991.

\bibitem{alkauskas2016tutorial}
Audrius Alkauskas, Matthew~D McCluskey, and Chris~G Van~de Walle.
\newblock Tutorial: Defects in semiconductors—combining experiment and
  theory.
\newblock {\em J. Appl. Phys.}, 119(18):181101, 2016.

\bibitem{kolobov1996origin}
AV~Kolobov.
\newblock On the origin of \textit{p}-type conductivity in amorphous
  chalcogenides.
\newblock {\em J. Non-Cryst. Solids}, 198:728--731, 1996.

\bibitem{abdullaev1965effect}
GB~Abdullaev, SI~Mekhtieva, D~Sh.~Abdinov, and GM~Aliev.
\newblock Effect of oxygen on some electrical properties of selenium.
\newblock {\em pss (b)}, 11(2):891--898, 1965.

\bibitem{freysoldt2014first}
Christoph Freysoldt, Blazej Grabowski, Tilmann Hickel, J{\"o}rg Neugebauer,
  Georg Kresse, Anderson Janotti, and Chris~G Van~de Walle.
\newblock First-principles calculations for point defects in solids.
\newblock {\em Rev. Mod. Phys.}, 86(1):253, 2014.

\bibitem{buckeridge2019equilibrium}
John Buckeridge.
\newblock Equilibrium point defect and charge carrier concentrations in a
  material determined through calculation of the self-consistent {F}ermi
  energy.
\newblock {\em Comput. Phys. Commun.}, 244:329--342, 2019.

\bibitem{bertoldo2022quantum}
Fabian Bertoldo, Sajid Ali, Simone Manti, and Kristian~S Thygesen.
\newblock Quantum point defects in 2{D} materials-the {QPOD} database.
\newblock {\em npj Comput. Mater.}, 8(1):1--16, 2022.

\bibitem{kangsabanik2022indirect}
Jiban Kangsabanik, Mark~Kamper Svendsen, Alireza Taghizadeh, Andrea Crovetto,
  and Kristian~S Thygesen.
\newblock Indirect band gap semiconductors for thin-film photovoltaics:
  High-throughput calculation of phonon-assisted absorption.
\newblock {\em Journal of the American Chemical Society}, 144(43):19872--19883,
  2022.

\bibitem{shockley1952statistics}
WTRW Shockley and WT~Read~Jr.
\newblock Statistics of the recombinations of holes and electrons.
\newblock {\em Physical review}, 87(5):835, 1952.

\bibitem{hall1952electron}
Re~N Hall.
\newblock Electron-hole recombination in germanium.
\newblock {\em Physical review}, 87(2):387, 1952.

\bibitem{kavanagh2021rapid}
Se{\'a}n~R Kavanagh, Aron Walsh, and David~O Scanlon.
\newblock Rapid recombination by cadmium vacancies in cdte.
\newblock {\em ACS Energy Letters}, 6(4):1392--1398, 2021.

\bibitem{kim2020upper}
Sunghyun Kim, Jos{\'e}~A M{\'a}rquez, Thomas Unold, and Aron Walsh.
\newblock Upper limit to the photovoltaic efficiency of imperfect crystals from
  first principles.
\newblock {\em Energy \& Environmental Science}, 13(5):1481--1491, 2020.

\bibitem{alkauskas2014first}
Audrius Alkauskas, Qimin Yan, and Chris~G Van~de Walle.
\newblock First-principles theory of nonradiative carrier capture via
  multiphonon emission.
\newblock {\em Physical Review B}, 90(7):075202, 2014.

\bibitem{sah1958electron}
Chih-Tang Sah and William Shockley.
\newblock Electron-hole recombination statistics in semiconductors through
  flaws with many charge conditions.
\newblock {\em Physical Review}, 109(4):1103, 1958.

\bibitem{alkauskas2016role}
Audrius Alkauskas, Cyrus~E Dreyer, John~L Lyons, and Chris~G Van~de Walle.
\newblock Role of excited states in shockley-read-hall recombination in
  wide-band-gap semiconductors.
\newblock {\em Physical Review B}, 93(20):201304(R), 2016.

\end{thebibliography}

\end{document}